\documentstyle[pramana,epsf,floats]{ias}

\def\asigma{\stackrel{\leftrightarrow}{\sigma}}

\def\gdot{\dot{\gamma}}

\def\ls{\lesssim}
\def\be{\begin{equation}}
\def\en{\end{equation}}                  
\def\p{\partial} 
\newcommand{\bi}[1]{\mbox{\boldmath$#1$}}
\newcommand{\av}[1]{\langle{#1}\rangle}

\def\bea{\begin{eqnarray}}
\def\ena{\end{eqnarray}}
\newcommand{\ppp}[3]{{\bigg(}\frac{\delta {#1}}{\delta {#2}}{\bigg )}_{#3}}
\renewcommand{\theequation}{\arabic{section}.\arabic{equation}}
\begin{document}
\title{Sheared Solid Materials}

\author{Akira  Onuki,  Akira Furukawa, and Akihiko Minami}
\address{Department of Physics, Kyoto University, Kyoto 606-8502, 
Japan}
\abstract{
We   present a 
time-dependent Ginzburg-Landau  model  of 
nonlinear elasticity in solid materials. 
We assume that   the elastic energy density is a 
periodic function of  the 
shear and tetragonal strains owing to 
the underlying lattice structure. 
With this new ingredient,  solving the equations 
yields  formation of dislocation dipoles or slips. 
In    plastic flow 
high-density dislocations emerge  
at large strains  to accumulate and 
grow into shear bands where the strains are 
localized.  In addition to 
the elastic displacement, 
 we also introduce 
the local free volume  {\it m}. 
For very small $m$ 
the defect  structures 
are metastable and long-lived   
where the dislocations are pinned  by 
the Peierls potential barrier. 
However, if the shear modulus decreases  
with increasing {\it m}, 
 accumulation of {\it m} 
around dislocation cores  eventually breaks  the Peierls potential 
leading to   slow relaxations  
in the stress and the free  energy (aging).  
As another application of our scheme, 
we also study dislocation formation  in 
two-phase alloys  (coherency loss) under shear strains, 
where   dislocations glide preferentially  
in the softer regions  and are trapped at the interfaces.  
}

\maketitle


\section{Introduction}

Nonequilibrium states  under shear 
have been extensively 
studied for  near-critical fluids 
and for various complex fluids including polymers, 
liquid crystals, surfactant systems, colloidal suspensions 
and so on \cite{Onukibook}. 
In  the soft condensed matter physics, 
considerable   attention has also 
been paid to jamming  rheology 
observed in sheared states 
of  supercooled liquids, 
soft glassy materials  such as 
dense microemulsions or granular materials 
 \cite{Liu_book}. 
 In  these systems
while  the particle size ranges from microscopic 
to macroscopic lengths,  
universal constrained dynamics is realized 
 under external shear forces. We mention that 
 mesoscopic dynamic heterogeneity and strong shear-thinning 
behavior have been observed in 
supercooled liquids (at relatively 
high $T$) \cite{Takeuchi,Muranaka,Yamamoto} and dense 
microemulsions  (at effectively 
low $T$) \cite{Okuzono}. 
In engineering, on the other hand, 
plastic flow  has long been studied 
in crystalline and amorphous 
 solids and in glassy polymers 
\cite{Cottrell,Nabarro,Spaepen_review1,Argon_review}.  
In crystals irreversible motions of dislocations 
give rise to plastic deformations  
and large strains produce high-density dislocations.  
In amorphous solids shear strains tend to be localized 
in narrow shear bands in plastic flow above 
a yield stress. The  width of 
such shear bands is microscopic  in the initial stage 
but can grow to mesoscopic  sizes, 
sometimes resulting in fracture.  Shear bands 
were numerically realized at large shear strains 
in  molecular dynamics  
simulations of  two-dimensional (2D) two-component 
glasses \cite{Deng,Falk} and in  simulations  
of a 2D  phenomenological model \cite{Bulatov}.   
Shear bands were  observed  also 
in  granular materials (at effectively zero temperature)
\cite{granular}.

We also stress 
relevance of dislocations in various  phase transformations 
in solids.  
Since the first work by Cahn \cite{Cahn1} 
most theoretical studies 
have been focused on the  coherent case in which 
the lattice planes are continuous through the 
interfaces \cite{Onukibook,Khabook,Fratzl}.   In the incoherent case   
 dislocations appear around the interfaces 
and the continuity is lost 
partially or even 
completely. Such incoherent  microstructures emerge 
  in various alloys  when  
 the lattice constants or 
the  crystalline structures 
of the two phases are not  close 
\cite{Strudel,loss}. 
Moreover, they are produced  
in plastic flow because  dislocations 
generated tend to be trapped  at the interfaces \cite{pollock}. 
Mechanical properties of two-phase solids are hence   
very different from those of one-phase solids.  
Simulation studies in this area  
are still at the beginning \cite{Desai,Chen,Sro}.

In this  paper we will present 
our recent nonlinear elasticity theory 
to understand  these diverse effects 
 in 2D \cite{OnukiJ,OnukiPRE,Minami}. 
Our  approach  is phenomenological, 
but its  merit over microscopic molecular dynamics simulations 
is that we can  put 
emphasis on any  aspects of the phenomena 
by  controlling   the parameters or changing  
the model itself.  
In Section 2 we will present 
a first version of our dynamic model, 
where the gross variables are 
the elastic displacement $\bi u$ and its velocity 
${\bi v}_{\rm L}= 
\p {\bi u}/\p t$. 
The stress-strain relation and emergence of plastic flow 
at large strains will be discussed. 
Metastability of the resultant 
defect structures will be 
ascribed to  the Peierls potential \cite{Nabarro,P}. In Section 3 
we will  introduce a free-volume field $m$   
 \cite{Cohen} into our  dynamic model, 
which will drastically change the dynamics 
of dislocations on long time scales.  As a result, we can gain  
 insights into the physical 
mechanism of aging in amorphous solids 
\cite{Struik,Kob,Sti,Lacks,Tar}. 
We expect that $m$ is a key order parameter 
for amorphous solids or glasses. 
In Section 4 we  will include the composition difference 
 $\psi=c_A-c_B$ for  binary  alloys 
to numerically realize dislocation networks 
in two-phase states under shear strain. 
Such complex effects have rarely been studied 
in physics but are of great technological importance.

\section{Mechanical  model}
\setcounter{equation}{0}

\subsection{Nonlinear elastic energy}
\begin{figure}[t]
\epsfxsize=6.25 in 
\centerline{\epsfbox{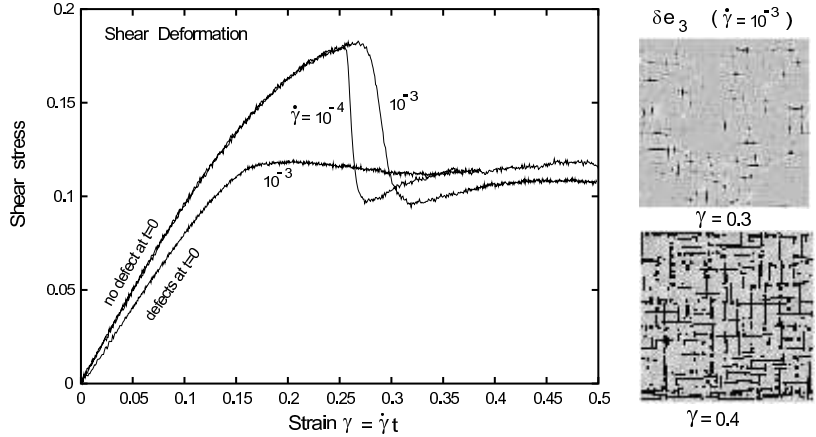}}
\caption{\protect
Stress-strain curves  from (2.6).  
Those at $\gdot=10^{-4}$ and $10^{-3}$ 
without defects at $t=0$ exhibit 
pronounced overshoot. 
Snapshots of 
 $\delta e_3=e_3-\gamma$  are given 
at  $\gamma=0.3$  and  $0.4$ (right).
 No marked overshoot 
appears in the  other curve  at  
$\gdot=10^{-3}$, where high-density dislocations 
are initially present.}
\label{1}
\end{figure}

We present  a  nonlinear elasticity model 
in 2D \cite{OnukiJ,OnukiPRE}. In terms of  the displacement 
 ${\bi u}=(u_x,u_y)$ from a reference crystal state, 
we   define the 
strain components as 
\bea 
e_1&=&\nabla_x u_x+ 
\nabla_yu_y , \nonumber\\
e_2&=& \nabla_x u_x- 
\nabla_yu_y, \nonumber\\
e_3&=& \nabla_x u_y 
+\nabla_yu_x, 
\label{eq:2.1}
\ena 
where $\nabla_x=\p/\p x$ and 
$\nabla_y=\p/\p y$.  We call $e_1$ the dilation strain, 
$e_2$ the tetragonal strain, and $e_3$ the shear strain. 
If we suppose a 2D triangular (or square)
 lattice,  the elastic energy should be invariant with 
respect to the rotations  of the reference frame 
by $\pm n\pi/3$ (or $\pm n\pi/2$) ($n=1,2, \cdots$). 
Under rotation of the reference frame 
by $\theta$, the shear strains $e_2$ and $e_3$ are changed 
to
\be  
e_2'= e_2 \cos 2\theta + e_3 \sin 2\theta, 
\quad e_3'= e_3 \cos 2\theta -e_2 \sin 2\theta.
\label{eq:2.2}
\en   
The  elastic energy is written as $F 
= \int d{\bi r} f_{\rm el} $
with the   density in  the form, 
\be 
f_{\rm el}= \frac{1}{2}{K}e_1^2 + {\Phi}(e_2, e_3),  
\label{eq:2.3}
\en 
where $K$ is the bulk modulus 
assumed to be a constant 
and $\Phi$ is the shear deformation energy. 
For a triangular lattice the simplest form of 
 ${\Phi}$ is given by 
\be 
{ \Phi} =  \frac{\mu}{6\pi^2} 
\bigg [ 3- \cos\pi(\sqrt{3}e_3-e_2) 
 - \cos\pi(\sqrt{3}e_3+e_2)
-  \cos(2\pi e_2) \bigg ], 
\label{eq:2.4}
\en  
which is invariant with 
respect to the rotation by $\pi/6$  in (2.2),  
is a periodic function of $e_3$  
with period $2/\sqrt{3}$ for $e_2=0$ 
(simple shear deformation),  and 
becomes the usual form 
$\mu (e_2^2+e_3^2)/2$ in  linear 
elasticity for small strains. 
A characteristic feature is that 
$\Phi$ in (2.4) is highly isotropic 
around its minima 
 in the $e_2$-$e_3$ plane. That is, 
around the reference state $e_2=e_3=0$, 
$\Phi$ may be teated as a function of 
$e=(e_2^2+e_3^2)^{1/2}$ only for $e \ls 1/2$.  
For  a square lattice structure  we  
have recently used 
\cite{Minami}
\be
{\Phi}=\frac{\mu_2}{4\pi^2}[ 1 - \cos (2 \pi e_2)] 
+\frac{\mu_3}{4\pi^2} [ 1 - \cos (2 \pi e_3)]   ,  
\label{eq:2.5}
\en   
which becomes $(\mu_2 e_2^2+\mu_3e_3^2)/2$ for 
small strains. 
One of the principal crystal axes 
is parallel to the $x$ axis in (2.4), 
while it is  along  or make an angle
of $\pi/4$ with respect to the $x$ axis in (2.5). 
If we allow the system to  rotate 
as a whole,   $f_{\rm el}$  becomes dependent  on 
the rotational strain 
$\omega= \nabla_xu_y-\nabla_yu_x$. 
In this sense our model is still incomplete and 
 further  generalization is  needed,  
particularly  in 
describing  polycrystalline states or melting phenomena.

\subsection{Dynamics}
\begin{figure}[t]
\epsfxsize=4 in 
\centerline{\epsfbox{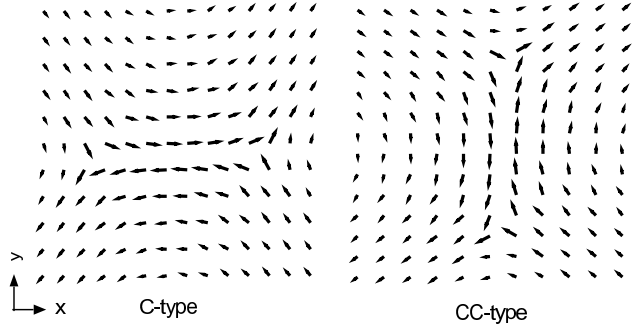}}
\caption{\protect
Displacement  $\bi u$ around 
type C and type CC slips  in the most favorable 
orientation in shear strain. 
The arrows are from the original 
undeformed  position 
to the displaced position.  }
\label{2}
\end{figure}

We assume that the lattice velocity 
${\bi v}_{\rm L} = \p{\bi u}/\p t$  
obeys 
\be
\rho \frac{\p}{\p t}{\bi v}_{\rm L} =  \nabla\cdot\asigma + 
 +\eta_0 \nabla^2{\bi v}_{\rm L}  + \nabla \cdot \asigma_{\rm R}  ,    
\label{eq:2.6}
\en 
where the stress tensor $\asigma= 
\{\sigma_{ij}\}$ is given by 
\be
\sigma_{xx} = K e_1 + \frac{\p \Phi}{\p e_2},\quad 
\sigma_{yy} = K e_1 - \frac{\p \Phi}{\p e_2},\quad 
\sigma_{xy} = \sigma_{yx} =
\frac{\p \Phi}{\p e_3}. 
\label{eq:2.7}
\en
It follows  the relation 
$\nabla\cdot\asigma= -{\delta F}/{\delta {\bi u}}$.  
The mechanical equilibrium condition 
$\nabla\cdot\asigma={\bi 0}$ 
is  equivalent to the extremum condition 
${\delta F}/{\delta {\bi u}}={\bi 0}$. 
We  introduce  the  viscosity  $\eta_0$ and 
the  random stress tensor $\asigma_{\rm R}$ \cite{random}.  
We integrated  (2.6) in 2D  
on a  $128\times 128$ square lattice with the 
 mesh size    equal to 
the lattice constant $a$. Use is made of  (2.4) (where  
a triangular lattice is supposed), but 
almost the same results  
follow  also using  $\Phi$ in (2.5). 
Here $\mu$ is assumed to be a constant $\mu_0$. 
Space and time 
are  measured in units of $a$ and 
$\tau_0= (\rho/\mu_0)^{1/2}a$. 
The  dimensionless   viscosity 
is   $\eta_0^* = \eta_0/\tau_0\mu_0$ 
and is  set equal to 1. 
Spatial derivatives are appropriately defined to 
yield  well-defined   microscopic 
slips.

 Fig.1 displays the stress-strain curves 
at constant shear rate  $\gdot$ applied for $t>0$ 
in units of $\mu_0$ and $\tau^{-1}$. 
For the  curves of $\gdot=10^{-3}$  and $10^{-4}$ 
with the pronounced  peak,   
we set   ${\bi u}={\bi 0}$ at $t=0$ 
supposing  a perfect crystal at $t=0$.  
They approach the elastic instability point 
$\gamma = \sqrt{3}/6$, where 
$\p \sigma_{xy}/\p \gamma=0$.  
Then the shear stress  drops sharply 
 with catastrophic slip formation. 
See the two snapshots of 
$\delta e_3=e_3-\gamma$. For the other curve at   
$\gdot=10^{-3}$ 
we initially put  
 high-density dislocations 
produced  after cyclic shearing \cite{OnukiPRE}.    
The stress overshoot is 
suppressed with increasing the 
initial disorder. 
In addition,  the slip spacing 
depends on  $\gdot$ in plastic 
flow.
For $\gdot=10^{-4}$  the spacing 
is a few times wider than 
in the case of $\gdot=10^{-3}$.

\subsection{Slips and Peierls potential}
\begin{figure}[t]
\epsfxsize=5in 
\centerline{\epsfbox{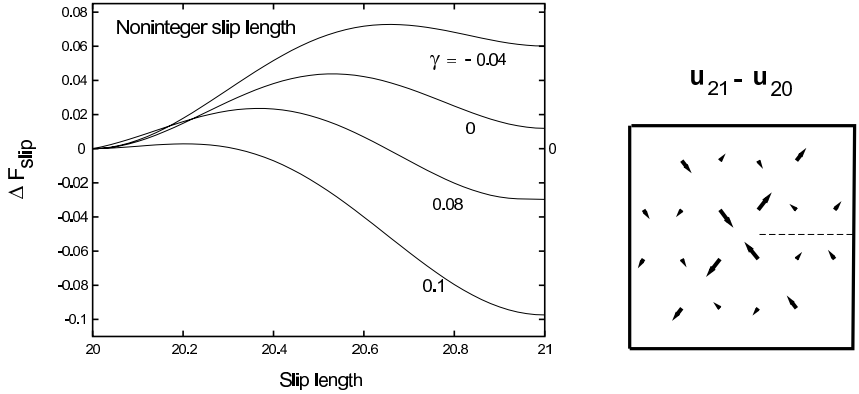}}
\caption{\protect
Slip energy difference 
$\Delta F_{\rm slip}(\ell) 
$ in the range $20\le \ell \le 21$ 
for $\gamma=-0.04, 0, 
0.08$, and $0.1$ in the model 
without free volume (left). 
Displacement difference  
${\bi u}_{21}-{\bi u}_{20}$ at $\gamma=0$ needed for 
 growth by unit length (right). 
 }
\label{3}
\end{figure}

\begin{figure}[t]
\epsfxsize=5in 
\centerline{\epsfbox{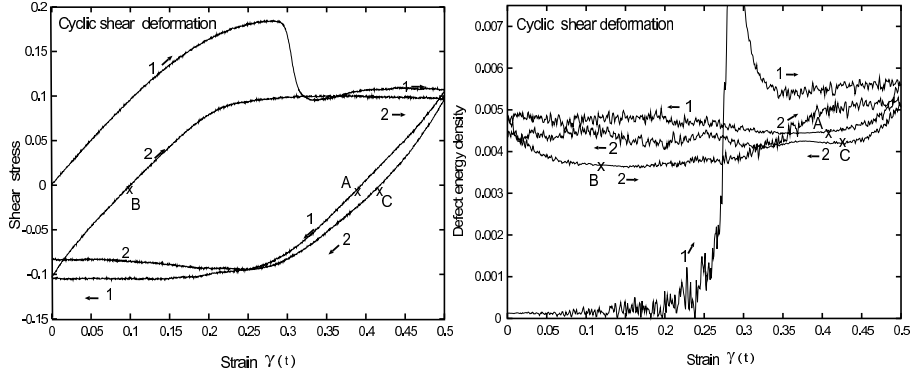}}
\caption{\protect
Shear stress 
$\av{\sigma_{xy}}$ (left) and 
defect energy density 
$f_{\rm D}$ defined by (2.12) (right)   
vs   strain $\gamma(t)$ 
for cyclic shear deformation  at  $\gdot=\pm 10^{-3}$.  
The shear stress vanishes at the three 
points A, B, and C ($\times)$.}
\label{4}
\end{figure}
 
A  fundamental flow  unit in our problem is a slip 
composed of a pair of edge dislocations.  
As shown in Fig.2, slips in 2D are divided into 
those of clockwise (C)  type 
and those of counterclockwise (CC)  type.
A slip  along one  of the crystal axes is  metastable 
without applied strains,  where 
 the slip length 
$\ell a$ is an integer multiple of the lattice 
constant $a$. It satisfies 
 the mechanical equilibrium 
condition and can be calculated numerically. 
For noninteger $\ell$ 
there arises an increase of the elastic energy, 
$U_{\rm P}(\ell)$, called the Peierls potential \cite{P}. 
Furthermore, under  applied strains,
$
\av{e_3} =\gamma$ and $\av{e_2} =\epsilon$, 
 the free energy  to create a slip 
 is given   by  \cite{OnukiPRE} 
\be 
 F_{\rm slip}
 = \frac{\ln \ell}{2\pi(1-\nu)}  
\mp   (\gamma \cos 2\varphi- \epsilon \sin 2\varphi) \ell 
+U_{\rm P}(\ell) ,  
\label{eq:2.8}
\en    
in units of $\mu a^2$ for the 
nearly isotropic $\Phi$ in (2.4).  Here  $\nu= (1-\mu/K)/2$, 
$\varphi$ is the angle of the slip with respect to the 
$x$ axis, and $U_{\rm P}(\ell)$ vanishes for integer $\ell$. 
In front of the second term, 
 $-$ is for  type C and $+$ is for type CC. 
For  simple shear deformation with $\gamma>0$ and $\epsilon=0$,  
 the most favorable slip orientation with the lowest 
$F_{\rm slip} $ is 
$\varphi=0$ for type C and $\varphi=\pi/2$ for type CC 
as in Fig.2.  For  uniaxial stretching 
 with $\gamma=0$ and $\epsilon>0$,  it is given by 
$\varphi=-\pi/4$ for type C and $\varphi=\pi/4$ for type CC.
In agreement with the last relations 
shear bands under uniaxial deformation 
have been observed to make  angles of $\pm \pi/4$ 
 with respect to the stretched direction 
   in  isotropic amorphous solids 
\cite{Argon_review,Deng,Falk,Bulatov} 
and   granular materials \cite{granular}.  
In real crystals slip orientations can be  
strongly influenced by crystal anisotropy.

For noninteger $\ell$,    
$U_{\rm P}(\ell)$ is path-dependent. 
Here  we calculate $F_{\rm slip}(\ell)$ 
in the range $20\le \ell\le 21$ 
for a type C slip 
parallel to the $x$ axis 
under  shear strain $\gamma$. Use 
is made of  the linearly extrapolated displacement,  
\be 
{\bi u}_\ell=  (1-\alpha){\bi u}_{20}+ 
\alpha {\bi u}_{21}, 
\label{eq:2.9}
\en 
where $\alpha=\ell-20$. The ${\bi u}_{20}$ and 
${\bi u}_{21}$ are numerically obtained 
for $\ell= 20$ and 21.      
Fig.3 shows  the free energy difference (left) 
\be 
\Delta F_{\rm slip}=F_{\rm slip}(\ell)-F_{\rm slip}(20), 
\label{eq:2.10}
\en  
which exhibits a maximum 
at an intermediate  $\ell$. 
The maximum 
approaches  20 as $\gamma \uparrow 
\gamma_{{\rm c}1}$ and 
21 as $\gamma \downarrow 
\gamma_{{\rm c}2}$, 
where $\gamma_{{\rm c}1}\sim 0.1$ 
and  $\gamma_{{\rm c}2}\sim -0.1$ here. 
Thus the slip is metastable for 
$\gamma_{{\rm c}2}<\gamma<\gamma_{{\rm c}1}$,  
but it  expands for $\gamma>\gamma_{{\rm c}1}$ 
and shrinks for $\gamma<\gamma_{{\rm c}2}$.
Fig.3 also indicates that 
about 10 particles  move  significantly 
when the  right end of the slip moves  by 
unit length (right). Against thermal agitations  
slips can be long-lived   if 
 the maximum of $\Delta F_{\rm slip}$ 
is much larger than $k_{\rm B}T$.

\subsection{Cyclic shear and defect energy density} 

Next
  we apply a cyclic shear deformation,  
where $\gdot(t)=10^{-3}$ in the time regions 
 $n t_{\rm p}<t<(n+1/2)t_{\rm p}$ 
and $\gdot(t)=-10^{-3}$ in the time regions $(n+1/2)
t_{\rm  p}<t<(n+1)t_{\rm p}$. 
For the first two cycles, 
Fig.4 shows the stress-strain curve (left)
and a  defect energy density $f_{\rm D}$ (right).  
The latter is defined as follows. 
We divide  
 the strain $\gamma(t)$ into 
an elastic strain $\gamma_{\rm el}$  
and a slip strain 
$
\gamma_{\rm s}= \gamma- 
\gamma_{\rm el}.
$ 
Roughly speaking, the elastic strain outside the 
slip lines gives rise  to the average stress, 
while the slip strain 
is caused  by the jumps of $u_x$ across the slips.
Regarding  the shear stress  in (2.7) 
as a function of $e_2$ and $e_3$, we 
  define $\gamma_{\rm el}$ by 
\be 
\sigma_{xy}(0,\gamma_{\rm el})  = \av{\sigma_{xy}}. 
\label{eq:2.11}
\en  
We define the  average 
defect energy density by 
\be 
f_{\rm D}= \av{f_{\rm el}}- \Phi(0,\gamma_{\rm el}),
\label{eq:2.12}
\en  
where $\Phi(e_2,e_3)$ is given by (2.4). 
The elastic energy density stored is then the  
sum of the regular elastic energy density 
and  the defect energy density. 
For small $\gamma_{\rm el}$   
we have $\gamma_{\rm el}  \cong 
 \av{\sigma_{xy}}/\mu_0$ and  
$\Phi \cong \mu_0 \gamma_{\rm el}^2/2$. 
In Fig.4  we notice 
(i)  residual strains  at  vanishing 
stress at points A,B, and C, (ii) no   overshoot
 in the stress  
 from  the second cycle, 
(iii) that $\av{\sigma_{xy}}<0$  at the ends of each  
cycle, and (iv) that $\av{\sigma_{xy}}(t)$  and $f_{\rm D}(t)$ 
are  roughly constant in
 plastic flow.

For $\gamma (t) \ls 0.3$ 
in the first cycle, however, 
 $f_{\rm D}(t)$   represents 
the elastic energy due to the inhomogeneous 
fluctuations of the local strains (mainly due to 
$\delta e_3$) with the peak height at 
0.012 (not shown in the figure). 
After this initial period,  
$f_{\rm D}(t)$ is 
 in a range of $0.004-0.005$ 
reasonably representing 
 the elastic energy due to the defects 
produced in plastic flow. 
Note that $f_{\rm D}(t)$ depends on $\gdot$ 
and it  is almost  constant 
around $0.002$ in plastic flow 
with $\gdot=\pm 10^{-4}$.

\section{Dynamic free-volume model}
\setcounter{equation}{0}

\subsection{Free volume} 
\begin{figure}[b]
\vspace{-4cm}
\epsfxsize=3.2in 
\centerline{\epsfbox{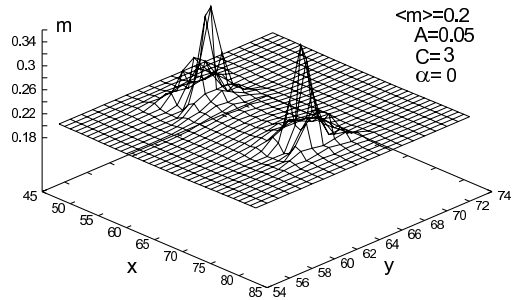}}
\caption{\protect
Free volume $m$ accumulated at the 
dislocation cores due to $m$-dependence 
of the shear modulus.  
 }
\label{5}
\end{figure}

In the presence 
of  point defects  or local free volume,   there  can be 
a  difference between 
the normalized density deviation 
$\delta\rho/\rho$ and $-e_1$  even in the linear elasticity. 
 Cohen, Flemming, and  Gibbs 
\cite{Flem} constructed a 
hydrodynamic description of solids including  
a  new variable which they called the vacancy field.  
Its  small deviation ($=\delta\rho/{\rho}+e_1$)  
was predicted to relax   diffusively (vacancy diffusion). 
More generally, 
since vacancies are point defects,  
let $m$ represent the local  free-volume fraction  
defined at each lattice point \cite{Cohen}, which can take 
continuous values. 
 Here we  introduce $m$ in our nonlinear elasticity scheme \cite{OnukiJ}.

As a reference state we suppose  a closely 
packed  crystal state with  $v_0$ 
being  a unit cell volume 
(or area in 2D).  We consider 
the cell number density $n_{\rm cell}$, 
 which changes with lattice dilation,  
and we    define $m$  as 
\be 
m= v_0(n_{\rm cell} - \rho/M), 
\label{eq:3.1}
\en  
where   $M$ is the mass in a unit volume of the 
reference state ($=$constant). 
We   assume that $m$ is non-negative ($m \ge 0$) 
neglecting excess packing such as  interstitials. 
If the density deviation $\delta\rho$ 
is small compared with the average density $\bar\rho$, $m$
 may be approximated as  
\be 
m= \av{m} - ( e_1+ \delta\rho/{\bar \rho}). 
\label{eq3:2}
\en 
The  average  free volume $\av{m}=v_0(\av{n_{\rm cell}} 
 - \bar{\rho}/M) $ 
decreases   with lowering  $T$ and/or 
increasing   $\bar \rho$ and 
represents the distance of the system from 
the closely packed perfect crystal. 
If the total volume of the system 
is fixed,  $m$  is  a conserved variable 
and $\av{m}$ is a constant independent of time. 
We shall see that  $\av{m}$ is an important parameter 
controlling slow relaxations of the defect structures. 
 
\subsection{Free energy and dynamic equations} 
\begin{figure}[t]
\hspace{4cm}
\epsfxsize=5.2in 
\centerline{\epsfbox{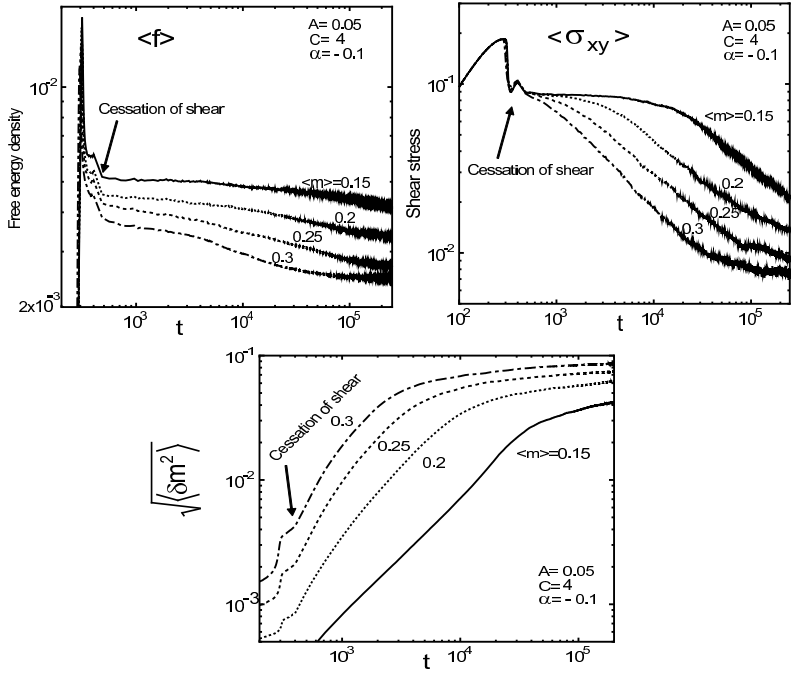}}
\caption{\protect
Free energy density 
$\av{f}(t)$, shear stress 
$\av{\sigma_{xy}}(t)$, and 
  free-volume variance $\sqrt{\av{\delta m^2}}(t)$ 
vs $t$. A shear flow  with 
 $\gdot=10^{-3}$ is applied in the time region 
$0<t<400$ and is stopped at $t=400$. 
Relaxations  become apparent with increasing 
the average free volume $\av{m}$. 
 }
\label{6}
\end{figure}
\begin{figure}[t]
\hspace{5.1cm}
\epsfxsize=5in 
\centerline{\epsfbox{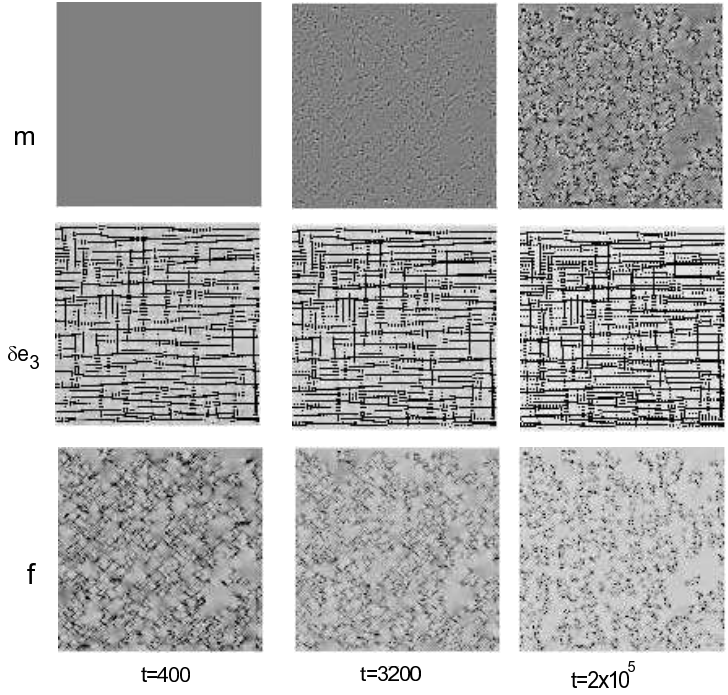}}
\caption{\protect
Snapshots of  free volume $m$, 
 shear strain deviation $\delta e_3$, 
and   free energy density $f$ 
at $t=400,3200$, and $2\times 10^5$ 
for $\av{m}=0.2$, 
 showing   
growth of heterogeneity of $m$, an increase of 
 plastic strain,  and a  decrease of $f$.  
  }
\label{7}
\end{figure}

In this work  we use  the following 
 free energy density,    
\be 
f= A m \ln (m/\av{m})  
    + \alpha m e_1+ 
\frac{1}{2}{K}e_1^2 +  {\Phi}(m,e_2, e_3).  
\label{eq:3.3}
\en 
The first term is an entropic term,  though 
its form  is not well justified.   However,  we may well 
replace it by the simpler form $A'm^2/2$ without loss 
of essential results of our simulations. 
The coupling parameter $\alpha$ 
is  assumed to be small 
since $|\delta m| \ll 
|\delta e_1|$ should hold  except close to the 
dislocation cores.  
We use the nearly isotropic  $\Phi$ in   
(2.4), but the shear modulus 
 $\mu$ decreases exponentially 
with increasing $m$ as  
\be 
\mu(m)= \mu_0 \exp(-C\delta m),
\label{eq:3.4}
\en 
where $\delta m=m-\av{m}$ 
and  $C$ is a positive 
constant. This relation  simply 
arises from the requirement  that solids should  soften 
with increasing the free volume. The same 
exponential form  was  used to describe 
 softening due to  point defects 
\cite{Granato}.   
If  $|\delta\rho|\ll \bar{\rho}$ 
and the strains are small, the 
 mass density deviation $\delta \rho$ obeys 
\be 
\frac{\p}{\p t} \delta\rho = -{\bar \rho} \nabla\cdot {\bi v}. 
\label{eq:3.5}
\en 
The mass velocity $\bi v$ is governed by   
\be
{\bar \rho} \frac{\p}{\p t}{\bi v} = 
 \nabla\cdot\asigma 
 +\eta_0 \nabla^2{\bi v} + \nabla \cdot \asigma_{\rm R}  .    
\label{eq:3.6}
\en 
Including the contribution from the pressure deviation 
$\bar{\rho}(\delta F/\delta \rho)_{\bi u}$ the 
force density  on the right hand side  is 
written in terms of  
$F=\int d{\bi r}f$ as 
\be 
\nabla\cdot\asigma= -{\bar \rho}\nabla\ppp{F}{\rho}{\bi u}
- \ppp{ F}{\bi u}{\rho}= - 
\ppp{ F}{\bi u}{m} ,  
\label{eq:3.7}
\en 
where use has been made of the identities  
 $(\delta/\delta {\bi u})_{\rho}= (\delta/\delta {\bi u})_{m}
- \nabla (\delta/\delta m)_{\bi u}$ 
and $(\delta/\delta m)_{\bi u}= -{\bar{\rho}} 
(\delta/\delta \rho)_{\bi u}$. The expressions for 
$\sigma_{ij}$ follow from (2.7) if 
$Ke_1$ is replaced by $Ke_1+\alpha m$ 
and the derivatives  are taken at fixed $m$.   
The lattice velocity 
$\p {\bi u}/\p t$ is different from 
the mass velocity ${\bi v}$ and is 
of the form \cite{Flem}, 
\be 
\frac{\p}{\p t}{\bi u} = {\bi v}_{\rm L} =  {\bi v} - L(m)  
\ppp{F}{\bi u}{\rho}  
+ {\bi \zeta}_{\rm R},   
\label{eq:3.8}
\en 
where $L(m)$ is the kinetic coefficient 
dependent on $m$ and 
${\bi \zeta}_{\rm R}$ is the random velocity 
 related to   $L(m)$
via the fluctuation-dissipation theorem. 
These dynamic equations are approximate. 
However, if the random forces are neglected,  
the total free energy $F_T=\int d{\bi r}[ f+\bar{\rho}
{\bi v}^2/2]$ decreases in time without external forces.
The system becomes steady only in mechanical equilibrium 
$\nabla\cdot\asigma={\bi 0}$.

From (3.5)-(3.8) we derive the equation for $m$,  
\be 
\frac{\p}{\p t} m=  \nabla\cdot L(m) 
\bigg [\nabla \ppp{F}{m}{\bi u} 
- \nabla\cdot\asigma \bigg ] + 
\nabla\cdot{\bi \zeta}_{\rm R}.  
\label{eq:3.9}
\en 
If the couplings to $\bi u$ are neglected, 
it follows  the nonlinear diffusion  equation 
$\p m/\p t= \nabla \cdot D(m)\nabla m$ 
with $D(m)= L(m) A/m$. 
Because the diffusion of $m$ should be 
drastically slowed down for small  
$m$, we 
assume the following form,  
\be 
D(m)=  D_0 \exp( -B/m),
\label{eq:3.10}
\en  
where $D_0$ and $B$ are positive constants. 
This form is also not well justified. 
It is worth noting that 
the same form   was used  for 
the relaxation rate of the free volume in granular 
matters \cite{deGennes}.  
Another aspect is the presence of 
$\nabla\cdot\asigma$ in the equation of $m$. 
A similar stress-diffusion coupling  is well-known 
in the dynamic equation of the composition 
for  gels and  viscoelastic fluid mixtures 
\cite{Onukibook,Toyo,Fre}. 
See the last section for more comments.

\subsection{Slow relaxation due to free-volume accumulation }

We choose 
$\bi v$, $\bi u$, and $m$ 
as fundamental variables 
and  solve (3.6),(3.8),  and (3.9) 
 by setting $A=0.05\mu_0$,  $B=1$, 
and $D_0= 2\times 10^{-3}$ \cite{scale}.
In Fig.5 we first show a steady-state profile of $m$ 
around a slip for $\av{m}=0.2$,  
$C=3$, and $\alpha=0$. 
Here the steady-state condition 
$\delta F/\delta m=$const. yields 
\be 
m \propto  \exp[  ( C \Phi-\alpha e_1)/A].   
\label{eq:3.11}
\en
Accumulation of  $m$  around  the dislocation cores 
becomes signficant for small $A$ and large $C$.  
Such heterogeneity  of $m$ should 
cause slow structural relaxations or aging, as 
expected in the literature \cite{Struik}.
As an example, a shear flow  at 
 rate $\gdot=10^{-3}$ is applied in the time region 
$0<t<400$ and is stopped at $t=400$.  
 Fig.6 displays  
time-evolution of the average free energy density 
$\av{f}(t)$,   the average stress $\av{\sigma_{xy}}(t)$, 
and  the variance 
$\sqrt{\av{\delta m^2}}(t)$. The last quantity 
represents  heterogeneity of 
$m$. The four curves 
are those for $\av{m}=0.15,0.2, 0.25$, and 0.3. 
The other parameter values  are given in the figure.  
The relaxations are 
more apparent with increasing 
$\av{m}$. We are adding  very small Langevin 
 noises  \cite{random}, so   noisy behavior appears  only 
at very long times. Therefore, 
the relaxations here are due to 
{\it deterministic}  diffusive accumulation of $m$ 
around randomly and densely distributed dislocations. 
If  the noise amplitude is increased, 
the relaxations become faster and noisy. 
\begin{figure}[t]
\epsfxsize=5.2in 
\centerline{\epsfbox{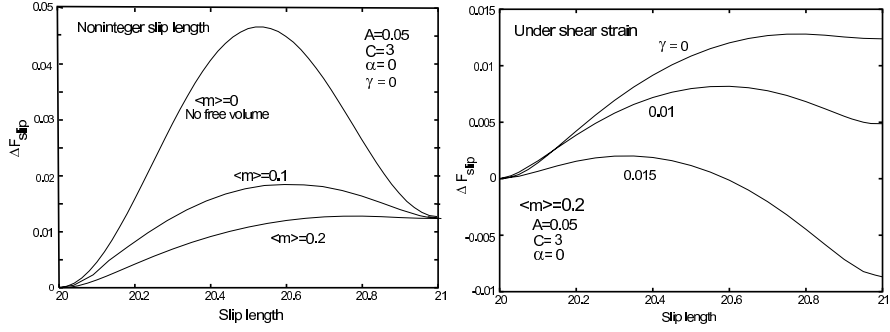}}
\caption{\protect
Slip energy difference $\Delta F_{\rm slip}(\ell) $ 
in the range $20\le \ell \le 21$ at $\gamma=0$ 
for $\av{m}=0,0.1$, and $0.2$  (right) 
and at $\av{m}=0.2$  for $\gamma=0,0.01$, and $0.015$ (left).  
The quasi-static condition 
$\delta F/\delta m=$const.is assumed along the path 
connecting the steady slip solutions with $\ell=20$ and 21. 
 }
\label{8}
\end{figure}

The origin of the slow relaxations 
may be   ascribed to weakening of 
the Peierls barrier  with 
accumulation of $m$ at the 
dislocation cores. In Fig.8 
  the slip energy 
$\Delta F_{\rm slip}$ defined  in (2.10) 
 is calculated in the range $20\le \ell \le 21$, 
where   $\bi u$ is fixed as in (2.9)  
but $m$ is relaxed until 
$\nabla\delta F/\delta m$ nearly 
vanishes at each $\alpha$. 
Namely,  $F$ is minimized  with respect to $m$ 
at    fixed $\bi u$.  
The slip-length change supposed here 
is {\it quasi-static}  taking place  
only on  time scales longer than 
$a^2/D(m)$. From 
Figs.3 and 8 we recognize  
that the maxima of $\Delta F_{\rm slip}$ 
much decrease  with increasing $\av{m}$  
and  that the  metastability  
is broken by applied  strains  much smaller 
than  in  the absence of   free volume.

\section{Two-phase alloys}
\setcounter{equation}{0}

\subsection{Composition coupled to nonlinear elasticity} 

We consider  a  binary alloy 
composed  of $A$ and $B$ atoms  
neglecting point defects. 
The compositions,  $c_A$ and $c_B$, 
of the two components satisfy  $c_A+c_B=1$. 
In real metallic alloys 
an order-disorder phase transition 
can also occur together with phase separation 
\cite{Onukibook,Khabook,Desai}.  
However, the composition difference 
$\psi= c_A-c_B$ is the  sole order parameter 
in this work. 
The free energy density $f$ is of the form, 
\be
f=  f_{\rm BW}(\psi) + \frac{C}{2} | \nabla \psi |^2
 +\alpha e_1\psi +   \frac{K}{2}  e_1^2 + 
{\Phi}(\psi, e_2, e_3 ).  
\en
The first term  is the Bragg-Williams 
free energy density\cite{Onukibook},   
\be
 f_{\rm BW}  
= \frac{k_{\rm B}T}{v_0}  
\bigg [ \frac{1 + \psi}{2}  \ln  ( 1 +
  \psi)  + \frac{1 - \psi}{2} 
\ln (1 - \psi)\bigg ]
 - \frac{k_{\rm B} T_{0}}{2v_0}\psi^2 ,  
\en
where $v_0$ is 
the volume of a unit cell, 
$ T_{0} $ is the mean-field critical temperature 
without  coupling to the elastic field. 
For  deep quenching this form is more appropriate than the 
 Landau expansion form. The second term in (4.1) 
is  the gradient free energy. 
There are two couplings between $\psi$ and the elastic field in $f$. 
First,   $\alpha$ is  the  strength of the coupling 
to $e_1$ arising from 
 a difference in the atomic sizes 
of  the two  species, leading to 
 {\it lattice misfit} in two-phase 
states. In particular, around a dislocation it gives rise to 
 a compositional 
Cottrell atmosphere \cite{Cottrell} in one-phase states and 
 preferential  nucleation \cite{CahnN} in the metastable 
 region. Second, 
 the shear deformation energy $\Phi$ is of 
the form of (2.5) with 
\be 
\mu_k  = \mu_{k 0} + \mu_{k 1} \psi 
\quad (k=2,3),   
\en 
where $\mu_{k 0}$ and $\mu_{k 1}$ 
are  constants. 
If $\mu_{21}>0$ and $\mu_{31}>0$, 
regions with larger (smaller) $\psi$ 
are harder (softer)   than  those 
with smaller (larger) $\psi$. 
The inhomogeneity 
in the shear moduli,  called {\it elastic inhomogeneity},  
plays a decisive role 
in the domain morphology 
in late stages,   because it 
gives  rise to   asymmetric 
elastic deformations in the two phases  
 \cite{Onukibook,Furukawa,Sagui}.

In the coherent case the 
mechanical equilibrium condition   
$ 
\nabla\cdot\asigma=  
 {\bi 0} 
$    may be 
assumed  even in dynamics. Here the  stress tensor 
$\asigma=\{\sigma_{ij}\}$  is 
 expressed as in (2.7). 
Using this condition in the linear elasticity,  
 the elastic field has been expressed in terms of $\psi$ 
in the previous theories 
\cite{Onukibook,Cahn1,Khabook,Fratzl}. 
We then find the following:\\
(i) The typical strain produced by the lattice misfit 
is given by 
\be 
e_0 = \alpha \Delta \psi/2L_0, 
\en 
where $\Delta\psi$ is the  difference 
of $\psi$ between the two phases and 
$ 
L_0(= K+ \mu_{20}
$ in 2D)    
is the longitudinal  modulus. 
In  the coherent condition  
we require 
 $e_0+\gamma \ls 1/4$ under 
 applied   strain $\gamma$.
(ii) In the weak limit of  cubic elasticity 
and  elastic inhomogeneity, 
one-phase states become linearly 
unstable for 
$T< T_{\rm s}/(1+ \av{\psi}^2/2)$,  
where 
\be 
T_{\rm s}= T_0+ v_0\alpha^2/L_0k_{\rm B} 
\en 
is the spinodal temperature 
of one-phase states at the critical composition.

In the incoherent case 
the mechanical equilibrium does not hold  
when dislocations are created 
and  are moving. 
We thus set up 
the dynamic equation (3.6) for  $\bi u$. 
Then the damping of $\bi u$ is governed by 
the kinetic viscosity $\eta_0/\rho$. 
The composition obeys  
\be 
\frac{\partial \psi}{\partial t} = 
\nabla \cdot  \lambda_0 ( 1 - \psi^2 )  \nabla
  \frac{\delta {F}}{\delta \psi}, 
\en 
where  $\bi u$ is fixed in  
$\delta F/\delta \psi$. We neglect Langevin noise terms 
in integrating these dynamic equations. 
In the dilute limit $c_A\rightarrow 0$, $c_A$ 
relaxes diffusively with the diffusion constant 
$ 
 D_0=\lambda_0 k_{\rm B}Tv_0^{-1}. 
$ 

\subsection{Dislocation formation} 
\begin{figure}[t]
\epsfxsize=5in 
\centerline{\epsfbox{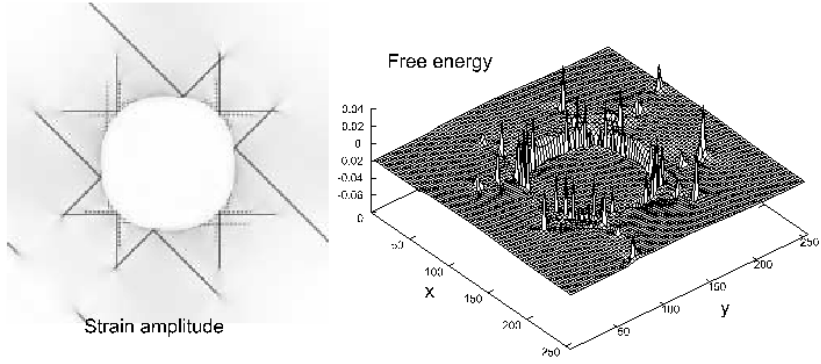}}
\caption{\protect
Strain amplitude 
$e$ (left) 
and   free energy density $f$ (right)  
 around an incoherent hard domain.  
Here $e$ is zero within  the hard domain and nonvanishing 
outside.  The peaks of $f$ are located at 
 the dislocation cores.  
 }
\label{9}
\end{figure}

\begin{figure}[b]
\epsfxsize=3.in 
\centerline{\epsfbox{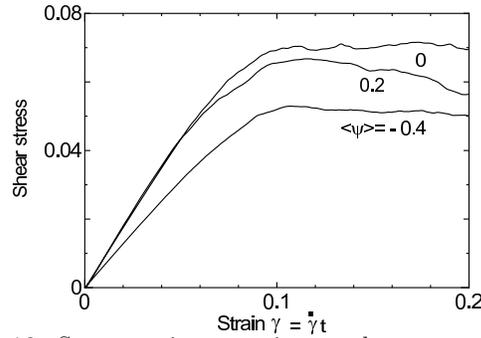}}
\caption{\protect
Stress-strain curves in two-phase states 
with $\av{\psi}= -0.4, 0$, and 0.2.  
There are no initial dislocations. 
Yield occurs with dislocation formation. 
 }
\label{10}
\end{figure}
\begin{figure}[t]
\epsfxsize=3.84in 
\centerline{\epsfbox{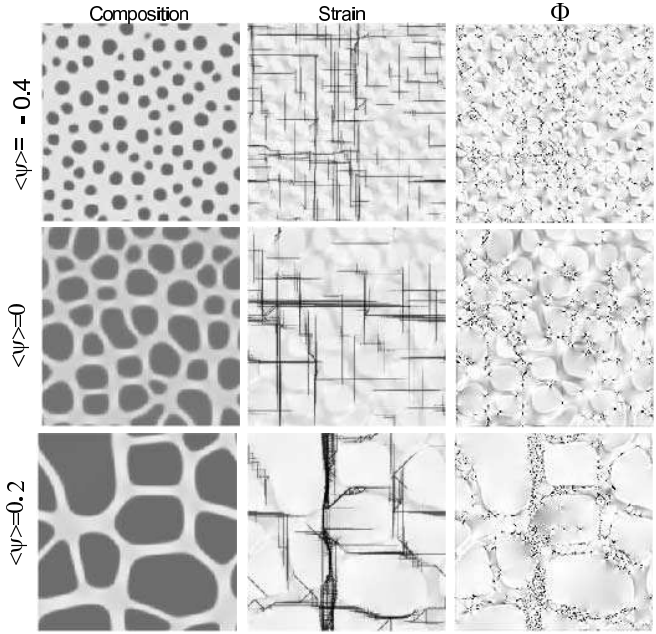}}
\caption{\protect
Composition difference $\psi$ (left), 
  shear strain deviation $\delta e_3=e_3-\gamma$ (middle), 
and  shear deformation energy $\Phi$ (right) in 
 plastic flow  at $\gamma=0.2$ in Fig.10.  
The average order parameter 
$\av{\psi}$ is -0.4, 0, and 0.2 from above. 
The black regions (right) represent harder  domains, 
which are enclosed by the softer phase (in gray). 
 }
\label{11}
\end{figure}

We integrate the dynamic equations  in 2D 
on a $256\times 256$ square lattice under the 
 periodic boundary condition by setting 
$K/\mu_{20}=4.5$, $\alpha/\mu_{20}=0.6$,    
$k_{\rm B}T_0/v_0 \mu_{20}=C/a^2\mu_{20}=
 0.05$, and $\mu_{21}=\mu_{31}=
0.6\mu_{20}$ \cite{Minami}. 
Space and time are  measured in  units 
of $a$ and  
$
\tau_0 = (\rho/\mu_{20})^{1/2}a,
$
respectively.  We set 
 the diffusion constant $D_0$ at 
$5\times 10^{-5}\eta_0/\rho$ 
and  the dimensionless viscosity 
$ \eta_0 /\tau_0 \mu_{20}$ at $0.1$. 
The time scale of $\psi$ is  much 
 longer than that of  the elastic field by 
four orders of magnitude.  In 
real solid alloys, these two time scales are 
even more  separated, 
probably  except for 
hydrogen-metal systems 
where the protons diffuse quickly 
\cite{Onukibook}.

Fig.9 shows  an incoherent 
 hard  domain 
obtained at deep quenching 
at  $T/T_0=1$. Slight cubic anisotropy 
$\mu_{30}/\mu_{20}=1.1$ is assumed, so 
the domain shape is a rounded square. 
We displays  the  strain amplitude 
$ 
e= ({e_2^2+e_3^2})^{1/2}$ (left) 
and the free energy density 
$f$ in (4.1) (right). 
The slips make  angles of $\pm \pi/4$ 
with respect to the $x$ axis 
where  $|e_2|$ is large, 
while they are parallel 
to the $x$ or $y$ axis  in the 
corner regions with large 
$|e_3|$. 
In  the interface region $f$   exhibits 
a cliff-like structure   arising from the 
gradient term  and  higher peaks  arising from 
 the dislocation cores.   Here we remark that 
dislocation clouds have been observed 
around $\gamma'$ precipitates 
of the  L1$_2$  structure in 
alloys with  large elastic misfit 
\cite{loss}.

Next, in the isotropic condition $\mu_{20}=\mu_{30}$,
 we  initially 
prepare coherent two-phase states at $T=2T_0<T_{\rm s}$ 
for three average compositions. We then 
apply  a shear flow at $\gdot = 10^{-3}$ for $t>0$. 
As can  be seen in the stress-strain curves in Fig.10,  
yield occurs at $\gamma \sim 0.1$ 
where dislocations start to be created. 
In Fig.11  slips glide in the softer regions 
and stop at the  interfaces. We can see that the 
black points in $\Phi$ representing dislocation cores 
are mostly trapped at the interfaces. 
These results are consistent with metallurgical 
experiments \cite{Strudel,pollock}.

\section{Summary and concluding remarks}

We have shown that the periodicity of 
the elastic energy density with respect to 
 $e_2$ and $e_3$ gives rise to 
proliferation of dislocations under large strains. 
Furthermore, we have found 
new approaches to 
unexplored complex effects 
with introduction of a new variable, 
the free volume or the composition, 
coupled to the elastic field. 
We give some  additional remarks.\\
(i) In our simulations slips emerge as  
long straight lines.  
If disorder is  fully introduced, 
 glide motions of slips in particular 
directions should be much limited. 
 In molecular dynamics (MD)
 simulations of two component glasses,  
the  degree of disorder should be sensitive to   
the size ratio of the two species.  When it was  rather close to 1, 
long  slips indeed  emerged \cite{Deng}.    
Performing   MD  simulations 
with various size ratios is 
thus  informative.\\ 
(ii) We mention the effect of 
elastic interaction  among   dislocations. 
In our case slips are more easily 
created around preexisting ones, 
leading  to  growth of shear bands, as already 
reported in Ref.\cite{Bulatov}.\\ 
(iii) In Section3, (3.6), (3.8), and (3.9) 
constitute a two-fluid model. We expect that   
 the fluctuations of 
$m$ and hence $\delta\rho$ can be   increased 
anisotropically under applied strains and after 
migration of $m$  \cite{OnukiJ}. 
This    effect  is 
intensified for small  $A$ in (3.3)  
and large $C$ in (3.4). 
In  polymers and gels  the composition fluctuations  
have  been observed to increase under strains  
 \cite{Onukibook,Fre}.\\ 
(iv) In Section 4, 
the composition has been  taken as a single 
order parameter. 
Extension  is needed 
to more general phase ordering  
 involving an 
order-disorder phase transition \cite{Onukibook,Khabook,Desai}
 or  diffusionless 
(Martensitic) structural phase 
transitions \cite{Onukibook,Khabook}.\\
(v) We conjecture that 
the origin of the dynamic heterogeneity observed in MD 
simulations \cite{Takeuchi,Muranaka,Yamamoto} could be ascribed to 
long-range correlations of the 
free volume.  This is because  the structural relaxation 
is caused by the free-volume redistribution 
around elastic inhomogeneity.\\
(vi) Finally we stress  that our nonlinear 
elasticity is still incomplete. 
For example, there should be 
the convective terms in 
(3.9) and (4.6). Also see the comment 
 below (2.5). 
Construction of a more complete 
theory is under way.





\begin{references}
\bibitem{random}  The dimensionless 
fluctuation-dissipation relation reads 
$\av{\sigma^{\rm R}_{ij}({\bi r},t)
\sigma^{\rm R}_{ij}({\bi r}',t')}=
2\epsilon_{\rm th} \eta_0^* 
\delta({\bi r}-{\bi r}')\delta(t-t')$ with 
$\epsilon_{\rm th}= k_{\rm B}T/\mu_0a^2$. 
We set $\epsilon_{\rm th}=4\times 10^{-5}$ 
in Fig.1  and $\epsilon_{\rm th}= 10^{-5}$ 
in Figs.6 and 7. However, $\epsilon_{\rm th}=0$ 
for the  other figures. 
For not very small Langevin noise 
the profiles  of the dynamic variables become 
fuzzy and we need to perform many simulation runs 
and take statistical averages.
\bibitem{scale} There is a freedom of scale change 
$m \rightarrow \lambda m$ and we may set 
$B=1$ without loss of generality. 
After this  scale change 
 $\delta\rho/{\bar \rho} = -\lambda\delta m -e_1$ 
determines the density deviation.
\end{references}
\end{document}